\documentclass[11pt,a4paper]{article}

\usepackage[english]{babel}

\usepackage[T1]{fontenc}
\usepackage[math]{iwona} 

\usepackage{amsmath,amssymb}
\usepackage{diffcoeff} 
\usepackage{graphicx}

\usepackage{subcaption}

\usepackage{placeins}

\usepackage{datetime}
\newdate{date}{17}{12}{2024}

\usepackage{natbib} 
\graphicspath{{./Fig/}}

\usepackage{tikz} \usetikzlibrary{shapes}

\usepackage{hyperref} 

\usepackage[capitalise]{cleveref}

\usepackage{calc}

\usepackage[paper=a4paper,portrait, tmargin=30mm, bmargin=30mm, rmargin=40mm, lmargin=25mm]{geometry} 

\usepackage{tabls}

\newlength\micolumna
\newcommand{\figwidth}{0.6\columnwidth}

\DeclareMathOperator*{\sign}{sign}

\title{Entry deterrence by exploiting economies of scope in data aggregation}

\author{Luis Guijarro, José Ramón Vidal, Vicent Pla\\
Universitat Politècnica de València, Spain\\
e-mail: lguijar@dcom.upv.es}

\date{}

\begin{document}

\thispagestyle{empty}

\begin{tabular*}{\textwidth}[t]{|p{\micolumna}|p{\micolumna}|}
	\hline
	\includegraphics[width=\micolumna,trim=2cm 17cm 2cm 2cm,clip]{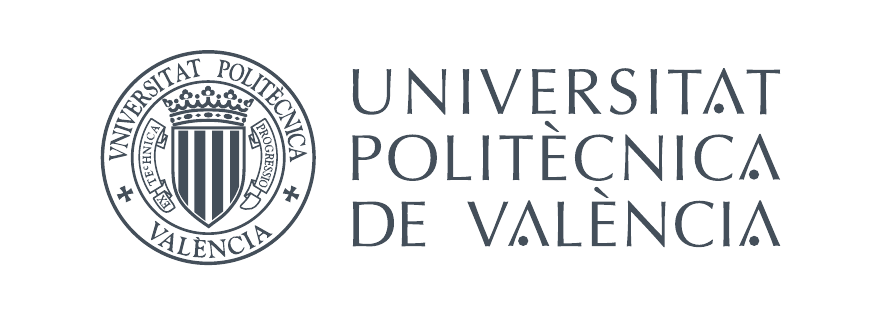}& 
	\raisebox{1.6cm}{\parbox{\micolumna}{Departamento de Comunicaciones}}\\
	\hline
	\multicolumn{2}{|c|}{}\\
	[0.20\textheight]
	\multicolumn{2}{|p{\textwidth-\tabcolsep*2}|}
	{\huge \begin{center}
Entry deterrence by exploiting economies of scope in data aggregation
\end{center}}\\
	\multicolumn{2}{|c|}{\huge (Document NETECON130E)}\\
	[0.20\textheight]

	\hline
	Authors: Luis~Guijarro,  & Outreach: Public\\ 
	José Ramón Vidal, Vicent Pla & \\
	\cline{2-2}
	 & Date: \displaydate{date} \\ 
	\cline{2-2}
	 & Version: a\\
	 \cline{2-2}
	 & Notebook: 130E2\\
	\hline
\end{tabular*}

\clearpage

\setcounter{page}{1}

\maketitle

\begin{abstract}
We model a market for data where an incumbent and a challenger compete for data from a producer. The incumbent has access to an exclusive data producer, and it uses this exclusive access, together with economies of scope in the aggregation of the data, as a strategy against the potential entry by the challenger. We assess the incumbent incentives to either deter or accommodate the entry of the challenger.
We show that the incumbent will accommodate when the exclusive access is costly and when the economies of scope are low, and it will blockade or deter otherwise. 
The results would justify an access regulation that incentivizes the entry of the challenger, e.g., by increasing production costs for the exclusive data. 
\end{abstract}

\section{Introduction}

There is widespread consensus that digital markets exhibit not only features that can increase competition relative to traditional markets, but also features that heighten concentration. Among the latter, the economies of scale and scope and the data advantage for incumbents are some of the most cited in both policy reports and academic works \citep{furman2019,cremer2019}.

We focus on a scenario where an incumbent enjoys both exclusive access to some data and also economies of scope in data aggregation. This may constitute an entry barrier and the incumbent, facing potential entry by a potential competitor, aka. challenger, will react strategically with a view either to make entry unprofitable or else to minimize the harm that entry causes. Entry is said to be deterred in the former case, and accommodated in the latter \citep{bain1956,belleflamme2015}.

This work analyses how an incumbent leverages the economies of scope in aggregating the data used as an input factor of its service provision, and when this leverage can be used to deter  a challenger. Our model is a duopsony, where there is a market which is used by two data aggregators in order to procure themselves with data.

The structure of the report is the following. Next section presents the related work. \cref{sec:model} describes model of the different agents in the market. \cref{sec:analysis} describes the strategic game that frames the decisions of the two aggregators. \cref{sec:results} discusses the numerical results obtained in a comparative statics analysis. And finally, \cref{sec:conclusions} puts forward some conclusions.

\section{Related work}\label{sec:related}

The theoretical analysis of the market for data between producers and aggregators is a flourishing area, where both personal data and machine data are studied. While the former area is focused on economic incentives and privacy concerns when consumers provide personal data\citep{dosis2023,economides2020,jones2020,ichihashi2021}, the latter focuses typically on the effects of the size and the combination of data sources. Our work belongs to the latter.

The previous works that are most related to this manuscript are \citet{calzolari2023} and \citet{carballa2023}, which model the market for data between producers and aggregators. This work differs from them in two respects: the model and the focus. Indeed, these two previous works model a monopolistic data aggregator, since the focus is on the incentives and the efficiency of the combination of data by the producers, while we model an aggregator incumbent and our focus is on the entry to the market by an aggregator challenger.

This manuscript assumes that economies of scale and scope are present in the data aggretation, which is an assumption that has been validated by the empirical literature. Among the vast amount of contributions in this area, we are indebted to \citet{carballa2022} and \citet{schaefer2023}.

\section{Model}\label{sec:model}

We model a scenario where there are two data aggregators (D1 and D2), and two data producers (P0 and P1).

Aggregator D1 is the incumbent; aggregator D2 is the challenger, who will bear an entry cost $F$ if it enters. Both aggregators D1 and D2 will compete in procuring themselves from data producer P1. Furthermore, aggregator D1 has exclusive access to data producer P0.

Data aggregation and processing is subject to economies of scale and economies of scope. The former are active up to a threshold amount, so that for amounts greater than the threshold diseconomies of scale take place. The latter are in effect when data from both P0 an P1 are aggregated, which implies that economies of scope can only be leveraged by D1.

\begin{figure}[!t]
	\begin{center}
	\begin{tikzpicture}[node distance=4.0cm,shorten >=1pt] 
	\node[ellipse,draw,align=center, minimum width=2cm](D1){D1};
	\node[ellipse,draw,align=center, minimum width=2cm](D2)[below of=D1]{D2};
	\node[circle,draw,align=center][right of=D1,yshift=1cm](P0){P0};
	\node[circle,draw,align=center][below of=P0](P1){P1};
	\path[->,blue]	(D1) edge [bend right] node[below]{$w_0$} (P0)
				(D1) edge [bend right] node[below]{$w$} (P1)
				(D2) edge [bend right] node[below]{$w$} (P1);
	\path[->,red,dashed]	(P0) edge [bend right] node[below]{$d_0$} (D1)
				(P1) edge [bend right] node[below]{$d_1$} (D1)
				(P1) edge [bend right] node[below]{$d_2$} (D2);
	\end{tikzpicture}
	\end{center}
	\caption{Data and Payment flow model.}\label{fig:payment}
\end{figure}
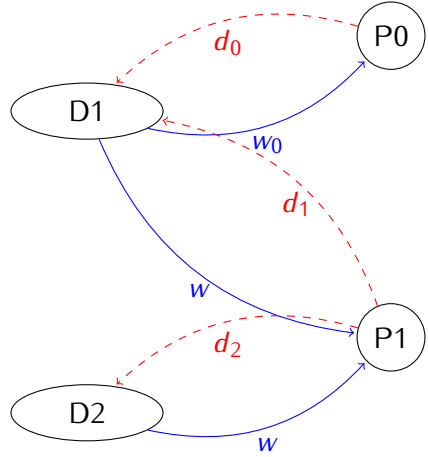

\cref{fig:payment} shows the data and payment flows between aggregators and producers.

\subsection{Data producers}

Data producers incur quadratic costs when generating data: $c d^2$ for producer P1 when producing amount $d$, and $c_0 d_0^2$ for producer P0 when producing amount $d_0$, and they are paid unit prices $w$ and $w_0$, respectively. They are assumed to be price takers, and therefore each one takes a production decision (either $d$ or $d_0$) in order to maximize profits, given prices.

Specifically, P1 profits are given by:
\begin{equation}
\Pi_1^p = w d - c d^2;
\end{equation}
and P0 profits are given by:
\begin{equation}
\Pi_0^p = w_0 d_0 - c_0 d_0^2.
\end{equation}
Profit maximization gives the following optimal data production decisions:
\begin{align}
w & = 2 c d \label{eq:P1_supply}\\
w_0 & = 2 c_0 d_0;\label{eq:P0_supply}
\end{align}
which are the inverse supply expressions, i.e., price as a function of the optimal quantity supplied or produced.

\subsection{Data aggregation}

We focus on data aggregation, that is, the accumulation of data items to a data set. This data set will feed a machine learning model that will provide a prediction tool to support the provision of a service.

Data aggregation may be subject to two kinds of efficiency gains: economies of scale and economies of scope. An intuitive way to distinguish between them is to use the analogy between a dataset and a two-dimensional spreadsheet. The number of columns would represent the number of variables, and the number of rows would represent the number of observations of the variables. When this dataset is used to make predictions, economies of scale improve prediction performance due to an increase in the number of rows, while economies of scope improve prediction performance due to an increase in the number of columns \citep{carballa2022}. 

As regards the economies of scale, we will consider the possibility that they run into diminishing returns beyond a threshold number of observations.

Based on the models proposed by~\citet{calzolari2023}, we model the value obtained from the aggregation of data by means of functions $\eta_A (d)$ and $\eta_O (d_1,d_2)$, which model the scAle effect and the scOpe effect, respectively.

\begin{figure}[t]
\begin{center}
\includegraphics[width=\figwidth]{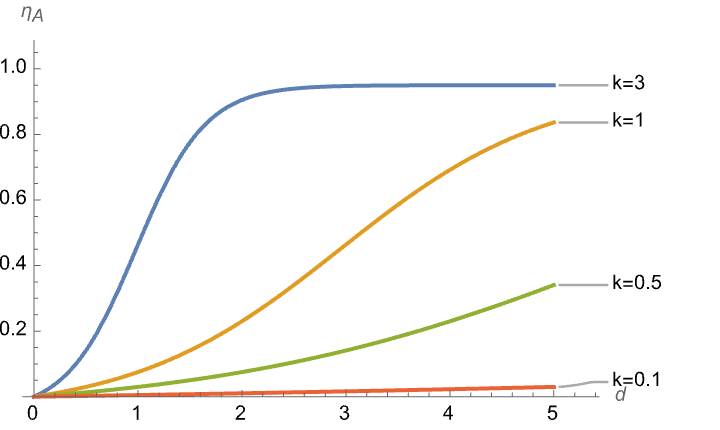}
\caption{Scale effect function}\label{fig:etaAPlot}
\end{center}
\end{figure}

The scale effect is modeled by a logistic function:
\begin{equation}\label{eq:scale}
\eta_A (d) = \eta_{max} \frac{1}{1+e^{-k(d-d_m)}},
\end{equation}
where $\eta_{max}$ is the maximum value attained, $k$ determines the slope, and $d_m$ determines the diminishing return threshold, and it is computed from parameter $\eta_0$ so that $\eta_A(0)=\eta_0$. \cref{fig:etaAPlot} shows $\eta_A$, for parameters $\eta_{max}=1$, $\eta_0=0.05$, and different values for $k$.

\begin{figure}[t]
\begin{center}
\includegraphics[width=\figwidth]{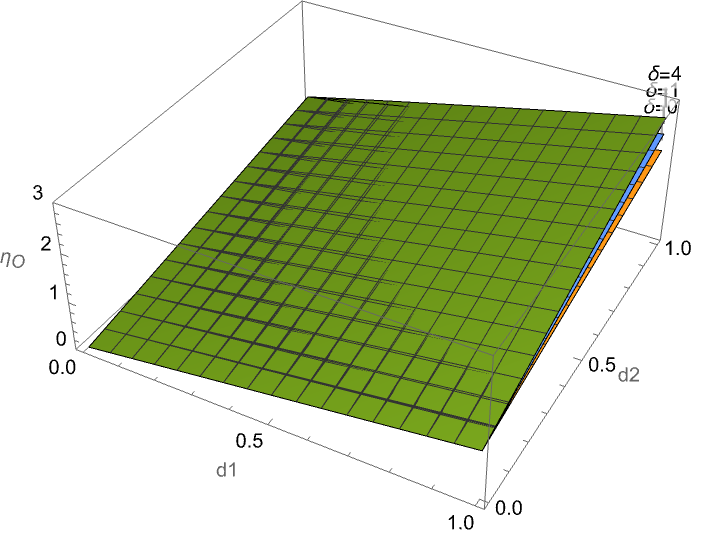}
\caption{Scope effect function}\label{fig:etaOPlot}
\end{center}
\end{figure}

\begin{figure}[t]
\begin{center}
\includegraphics[width=\figwidth]{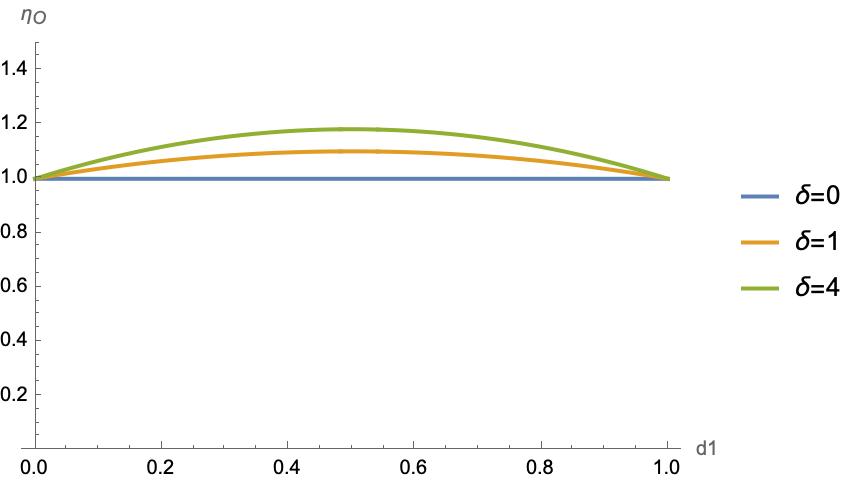}
\caption{Scope effect function when $d_1+d_2=1$}\label{fig:etaO2DPlot}
\end{center}
\end{figure}

The scope effect is modeled by a superadditive operation that for two contributions, $d_1$ and $d_2$, which is the case at hand, is given by:
\begin{equation}\label{eq:scope}
\eta_O (d_1,d_2) = \left( (d_1+1)^{\frac{1}{1+\delta}}+(d_2+1)^{\frac{1}{1+\delta}}      \right)^{(1+\delta)} -1.
\end{equation}
\cref{fig:etaOPlot} shows $\eta_O (d_1,d_2)$, for different values for $\delta$. \cref{fig:etaO2DPlot} shows $\eta_O (d_1,1-d_1)$, where the superadditivity effect is clearly appreciated.

\begin{figure}[t]
\begin{center}
\includegraphics[width=\figwidth]{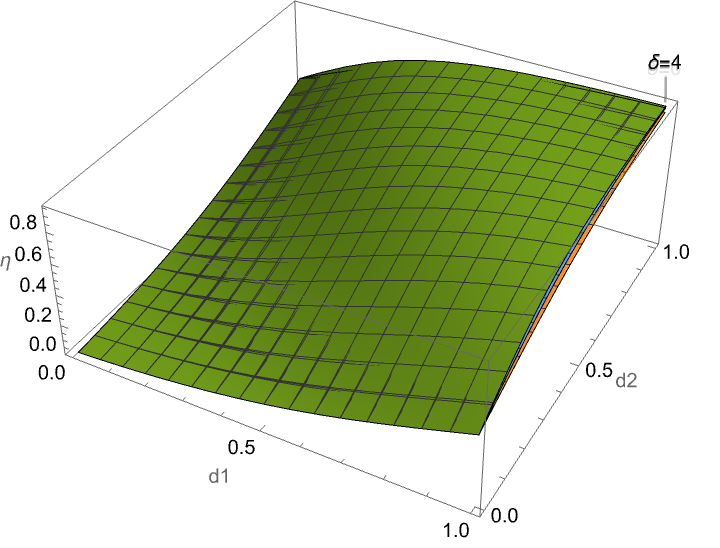}
\caption{Combined Scope and scale effects function}\label{fig:etaPlot}
\end{center}
\end{figure}

Finally, the combined effect $\eta_A (\eta_O (d_1,d_2))$ is plotted in \cref{fig:etaPlot}, for $\eta_{max}=1$, $\eta_0=0.05$, $k=3$ and different values for $\delta$.

We assume that aggregator revenues equal the value computed from $\eta_A (d)$ and $\eta_O (d_1,d_2)$, so that aggregator D1 will get $\eta_A (\eta_O (d_0,d_1))$ and aggregator D2 will get $\eta_A (d_2)$. We assume that the only relevant costs are the payments to the data producers, so that aggregator profits are given by:
\begin{align}
\Pi_1 & = \eta_A (\eta_O (d_0,d_1)) - w d_1 - w_0 d_0 \label{eq:D1_profits}\\
\Pi_2 & = \eta_A (d_2) - w d_2 - F. \label{eq:D2_profits}
\end{align}

\section{Analysis}\label{sec:analysis}

We model a strategic interaction between aggregators D1 and D2, whereby the following sequence of decisions are taken:
\begin{enumerate}
\item Stage 0: Incumbent aggregator D1 decides the amount $d_0$ to procure from producer P0.
\item Stage 1: Challenger aggregator D2 decides whether to enter or not to enter the data market supplied by producer P1.
\item Stage 2.a: If D2 has entered, D1 and D2 compete in quantities in procuring data from producer P1.
\item Stage 2.b: If D2 has not entered, D1 acts as a monopsonist and decides how much data to procure from P1.
\end{enumerate}

We assume that there is perfect recall in the sequential game described above, that is, both players D1 and D2 are aware of the decisions taken in previous stages by all players, and we look for Subgame Perfect Nash equilibria, which we derive using backward induction.

\subsection{Stage 2.b}

Given that decisions have been taken at Stages~0 and~1 and are known, so that $d_0$ and $w_0$ are known and D2 has not entered, D1 is now a monopsonist and will choose $d_1$ in order to maximize its profits:
\begin{align} 
 \underset{d_1}{\max} &\quad  \Pi_1 \\
 \text{subject to } &\quad   d_1 \geq 0 , \nonumber
\end{align} 
and subject to the anticipated fact that P1 will choose $d_1$ according to~\eqref{eq:P1_supply}, where $d=d_1$, so that
\begin{align}\label{eq:D1mon_PMP}
 \underset{d_1}{\max} & \quad \eta_A (\eta_O (d_0,d_1)) - (2 c d_1) d_1 - w_0 d_0 \\
 \text{subject to } & \quad  d_1 \geq 0. \nonumber
\end{align} 
The solution to~\eqref{eq:D1mon_PMP} will be denoted by $d_1^m$.

\subsection{Stage 2.a}

Given that decisions have been taken at Stages~0 and~1 and are known, so that $d_0$ and $w_0$ are known and D2 has entered, D1 and D2 will compete in data quantities $d_1$ and $d_2$ in order to maximize its profits, that is, Nash equilibrium $(d_1^*,d_2^*)$ will be the solution of the following simultaneous profit maximization problems, where both players anticipate that P1 will choose $d$, which matches $d_1+d_2$, according to~\eqref{eq:P1_supply}:
\begin{align} 
 \underset{d_1}{\max} & \quad \eta_A (\eta_O (d_0,d_1)) - 2 c (d_1+d_2) d_1 - w_0 d_0  \label{eq:D1duo_PMP}\\
 \text{subject to } & \quad  d_1 \geq 0. \nonumber\\ 
 \underset{d_2}{\max} & \quad \eta_A (d_2) - 2 c (d_1+d_2) d_2 - F  \label{eq:D2duo_PMP}\\
 \text{subject to } & \quad  d_2 \geq 0. \nonumber
\end{align} 

\subsection{Stage 1}

The decision at Stage~1 by aggregator D2 is whether to enter or not. It anticipates the results of Stage~2, and it decides so that profits $\Pi_2$ are maximized. If D2 enters, then it will get profits $\Pi_2(d_1^*,d_2^*)$, according to~\eqref{eq:D1duo_PMP} and~\eqref{eq:D2duo_PMP}. If D2 does not enter, then it will get zero profits. Thus, D2 will enter if and only if $\Pi_2(d_1^*,d_2^*) \geq 0$.

\subsection{Stage 0}

The decision at Stage~0 is taken by aggregator D1 and it is the data amount $d_0$ to procure from producer P0 and the unit price $w_0$. Since it anticipates that P0 will choose $d_0$ according to~\eqref{eq:P0_supply}, the decision boils down to $d_0$. D1 also anticipates the decision to be taken by D2 at Stage~1 and the two alternatives to Stage 2, i.e., Stage~2.a and Stage~2.b. And it will maximize its profits, that is,
\begin{align} \label{eq:D1_PMP}
 \underset{d_0}{\max} & \quad \eta_A (\eta_O (d_0,d_1)) - 2 c (d_1+d_2) d_1 - (2 c_0 d_0) d_0 \\
 \text{subject to } & \quad  d_0 \geq 0, \nonumber\\
 & \quad \text{Stage 1} \nonumber, \\ 
 & \quad \text{Stage 2}, \nonumber
\end{align} 
which can be decomposed in the following profit maximization subproblems:
\begin{itemize}

\item the one that maximizes D1 profits anticipating correctly that D2 will not enter and D1 will remain the monopsonist:

\begin{align} 
 \Pi_1^{deter}& =& \underset{d_0}{\max} & \quad \eta_A (\eta_O (d_0,d_1^m)) - (2 c d_1^m) d_1^m - (2 c_0 d_0) d_0 \label{eq:D1deter_PMP} \\
 & &\text{subject to } & \quad  d_0 \geq 0, \nonumber\\
 & & & \quad \Pi_2(d_1^*,d_2^*) \leq 0, \label{eq:deterconstraint} \\
 & & & \quad \text{where }  (d_1^*,d_2^*) \text{ is the solution to~\eqref{eq:D1duo_PMP}  and~\eqref{eq:D2duo_PMP}}, \nonumber \\
 & & & \quad d_1^m \text{ is the solution to~\eqref{eq:D1mon_PMP}}. \nonumber
\end{align} 
We say that the incumbent \emph{deters} the entrant. If \eqref{eq:deterconstraint} is not binding at the solution to~\eqref{eq:D1deter_PMP}, then the entry is said to be \emph{blockaded}, instead.

\item the one that maximizes D1 profits anticipating correctly that D2 will enter and compete:

\begin{align} \label{eq:D1accom_PMP}
 \Pi_1^{accommodate} &= & \underset{d_0}{\max} & \quad \eta_A (\eta_O (d_0,d_1^*)) - 2 c (d_1^*+d_2^*) d_1^* - (2 c_0 d_0) d_0 \\
 & & \text{subject to } & \quad  d_0 \geq 0, \nonumber\\
 & & & \quad \Pi_2(d_1^*,d_2^*) \geq 0, \nonumber\\
 & & & \quad (d_1^*,d_2^*) \text{ is the solution to~\eqref{eq:D1duo_PMP} and~\eqref{eq:D2duo_PMP}}. \nonumber
\end{align} 
We say that the incumbent \emph{accommodates} the entrant.

\end{itemize}
D1 will deter if $\Pi_1^{deter} \geq \Pi_1^{accommodate}$ and accommodate otherwise. 

Based on terminology coined by~\citet{bain1956}, and following~\citet{tirole1988} and~\citet{belleflamme2015}, we characterize below the strategic choice of $d_0$, in the entry deterrence and accommodation cases.

In the deterrence case, as stated above,
\begin{equation}
\Pi_2(d_1^*,d_2^*) = 0
\end{equation}
If we totally differentiate $\Pi_2$ with respect to $d_0$:
\begin{equation}
\diff{\Pi_2}{d_0}=\diffp{\Pi_2}{d_0}+\diffp{\Pi_2}{d_1}\diffp{d_1^*}{d_0}+\diffp{\Pi_2}{d_2}\diff{d_2^*}{d_0},
\end{equation}
where the fist term would be the direct effect, which is zero in our model, since $\Pi_2$ does not depend directly on $d_0$; the second term is the strategic effect (hereafter, SED), and the third term vanishes at the second-stage Nash equilibrium. Thus, the total effect of $d_0$ on $\Pi_2$ is the SED.

When $\difs{\Pi_2}{d_0}<0$ (or, equivalently, $SED<0$), we say that the procured amount $d_0$ (\emph{investment} in Bain's parlance) makes the incumbent \emph{tough}; and \emph{soft} otherwise. In the former, the incumbent has an incentive (compared to a counter-factual scenario where no strategic effect is present) to overinvest in order to push the entrant profit to zero, and this entry-deterrent strategy is known as ``top dog strategy''. In the latter, the incumbent has an incentive to underinvest in order to push the entrant profit to zero, and this entry-deterrent strategy is known as ``lean and hungry strategy''.

In the accommodation case, as stated above, the incumbent maximizes its own profit. If we totally differentiate $\Pi_1$ with respect to $d_0$:
\begin{equation}
\diff{\Pi_1}{d_0}=\diffp{\Pi_1}{d_0}+\diffp{\Pi_1}{d_1}\diffp{d_1^*}{d_0}+\diffp{\Pi_1}{d_2}\diffp{d_2^*}{d_0},
\end{equation}
where the fist term is the direct effect; the second term vanishes at the second-stage Nash equilibrium, and the third term is the strategic effect (hereafter, SEA). Assuming that the solution to~\eqref{eq:D1accom_PMP} is interior, i.e., $\difs{\Pi_1}{{d_0}}=0$,  the direct effect and the SEA have opposite signs and cancel out at the solution..

Assuming that the firms' second-stage choices have the same nature, i.e., $\difsp{\Pi_2}{d_1}$ and  $\difsp{\Pi_1}{d_2}$ have the same sign, the following relationships between SEA and SED can be derived:
\begin{equation}
\begin{split}
\sign{\text{SEA}}& =\sign{\diffp{\Pi_1}{d_2}\diffp{d_2^*}{d_0}}=\sign{\diffp{\Pi_2}{d_1}\diffp{d_2^*}{d_0}}\\
& = \sign{\diffp{\Pi_2}{d_1}\diffp{d_2^*}{d_1}\diffp{d_1^*}{d_0}}=\sign{\diffp{\Pi_2}{d_1}\diffp{d_1^*}{d_0}}\sign{\diffp{d_2^*}{d_1}}\\
& =\sign{\text{SED}} \times \sign{\diffp{d_2^*}{d_1}}
\end{split}
\end{equation}

We note that $\difsp{d_2^*}{d_1}$ is the slope of the entrant's best response. When both best responses' slopes are positive, we say that competitors are ``strategic complements'', since a competitor reacts in the same direction against a change on the opponent's strategy, e.g., increasing $d_2$ against an increase in $d_1$. When they are negative, we talk about ``strategic substitutes''.

Thus, when the competitors are strategic substitutes, the signs of the SEA and of the SED are different, so that, when the investment makes the incumbent tough (i.e., SED<0), then the SEA is positive and the incumbent has an incentive to overinvest, that is, it chooses the same ``top dog strategy'' as in the deterrence case, but now it consists in being big in order to look offensive and prompting a ``friendly'' reaction. Analogously, when  the investment makes the incumbent soft (i.e., SED>0), then the SEA is negative and the incumbent has an incentive to underinvest, that is, a ``lean and hungry strategy'', as in the deterrence case.

However, when the competitors are strategic complements, the sign of the SEA and of the SED are the same, so that, when the investment makes the incumbent tough (again, SED<0), then the SEA is now negative and the incumbent has an incentive to underinvest, that is, a different ``puppy dog strategy'', which consists of being weak or small to look inoffensive, so as to trigger a favourable response from the entrant. And when  the investment makes the incumbent soft (again, SED>0), then the SEA is positive and the incumbent has an incentive to overinvest, that is, a different ``fat cat strategy'', which consists of being big to also look inoffensive.

For all numerical results discussed in the next section, the slope of the best responses are negative, so that the competitors are of a strategic substitutable nature. And the SEA is positive. Therefore, the SED is negative. Thus, both deterrence and accommodation calls for overinvestment and, therefore, a ``top dog strategy''.

\section{Results and discussion}\label{sec:results}

We have solved the above maximization problems numerically, using software Wolfram Mathematica 13. Stage 2.b maximization problem~\eqref{eq:D1mon_PMP} has used an implementation of the Couenne library for solving mixed integer nonlinear optimization problems. Stage 2.a game problem~\eqref{eq:D1duo_PMP} and~\eqref{eq:D2duo_PMP} combines the above with an iterated algorithm for fixed point computation. Finally, Stage 0 maximization problem has used Differential Evolution, a direct search method for global optimization, for the upper level problem under deterrence~\eqref{eq:D1deter_PMP}, and the BFGS method, a Quasi-Newton algorithm for local optimization, under accommodation~\eqref{eq:D1accom_PMP}.

We discuss the numerical results through comparative statics, that is, we vary different parameters and assess how the equilibrium $(d_0^*,enter \lvert not enter,(d_1^*,d_2^*) \lvert d_1^m)$ varies. In order to evaluate the different equilibria, we compute the profits that aggregators D1 and D2 and producers P1 and P0 obtain, as well as the Social Welfare, which is defined as the sum of the profits of all agents.

The following values for the parameters are used, if not stated otherwise. For the data value functions $\eta_A$ and $\eta_O$, $\eta_{max} =1$, $\eta_0=0.05$, $k=3$, and $\delta=1$. Cost parameters are $c=2/3$ and $c_0=5/6$. Three fixed costs values are used: $F=0.00005, 0.0005, 0.0007$, referred to hereafter as low, medium and high, respectively.

\subsection{Comparative statics: \texorpdfstring{$c_0$}{c0}}

We vary $c_0$, which is producer P0's cost parameter, in the range $[ 3/6, 6/6 ]$. We evaluate therefore the effect of increasing the production cost of the data that is supplied under exclusivity to aggregator D1 and that is the source of the economies of scope that supports its competitive advantage.

\begin{figure}
\centering
\begin{subfigure}{.32\textwidth}
\includegraphics[width=\linewidth]{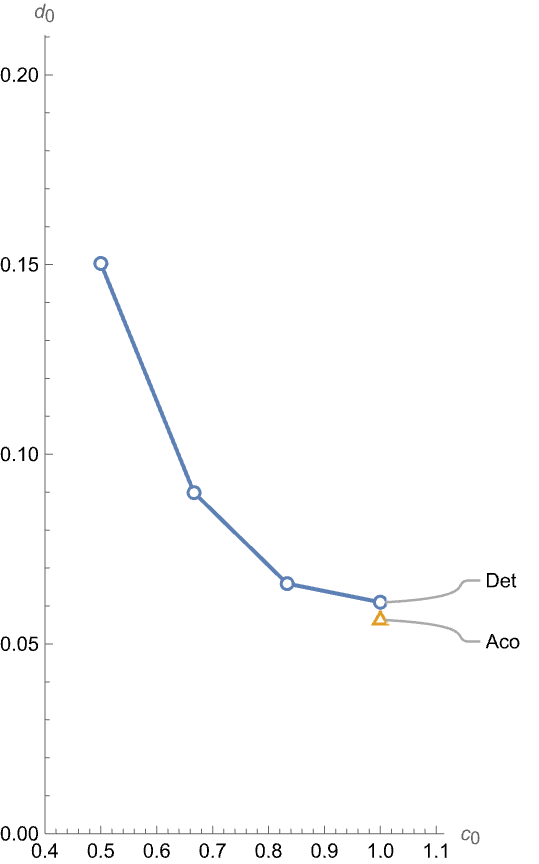}
\subcaption{High entry cost}\label{fig:d0c0HighFPlot}
\end{subfigure}
\begin{subfigure}{.32\textwidth}
\includegraphics[width=\linewidth]{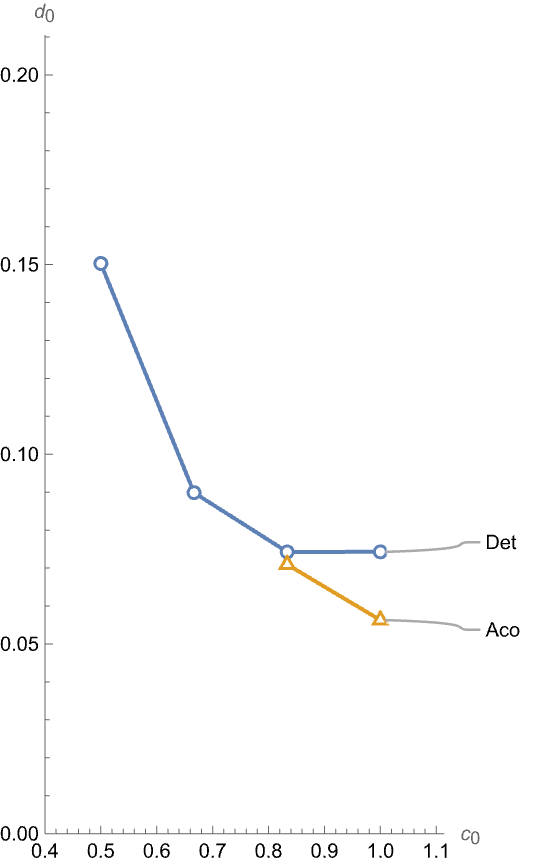}
\subcaption{Medium entry cost}\label{fig:d0c0MedFPlot}
\end{subfigure}
\begin{subfigure}{.32\textwidth}
\includegraphics[width=\linewidth]{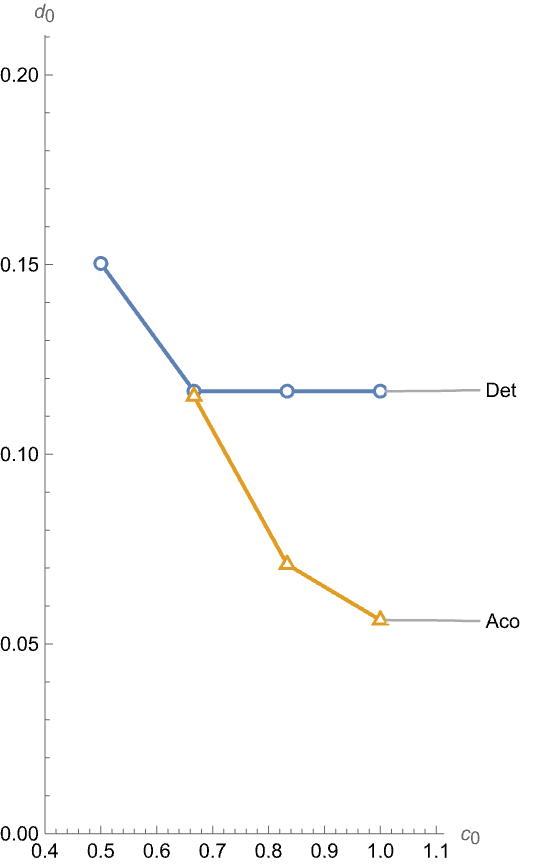}
\subcaption{Low entry cost}\label{fig:d0c0LowFPlot}
\end{subfigure}
\caption{Profit maximizing deterring and accommodating $d_0$ as a function of $c_0$.}
\end{figure}

\begin{figure}[t]
\begin{center}
\includegraphics[width=\figwidth]{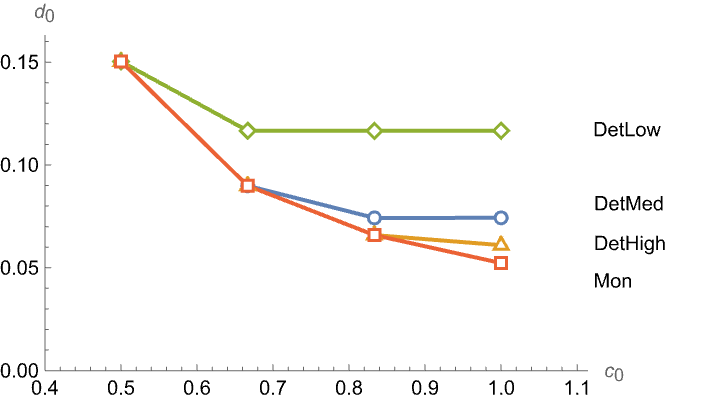}
\caption{Profit maximizing deterring and monopsonist $d_0$ as a function of $c_0$. Different entry costs.}\label{fig:d0Detc0Plot}
\end{center}
\end{figure}

\begin{figure}
\centering
\begin{subfigure}{.32\textwidth}
\includegraphics[width=\linewidth]{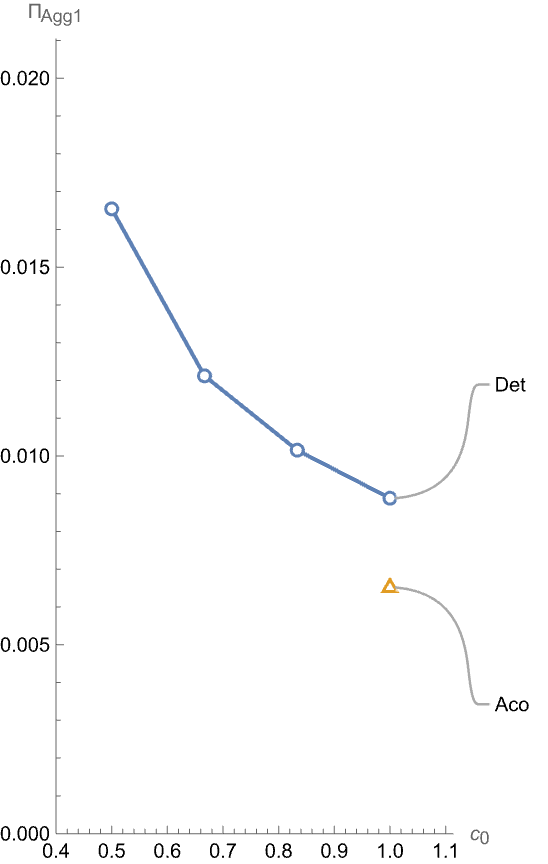}
\subcaption{High entry cost}\label{fig:ProfAgg1c0HighFPlot}
\end{subfigure}
\begin{subfigure}{.32\textwidth}
\includegraphics[width=\linewidth]{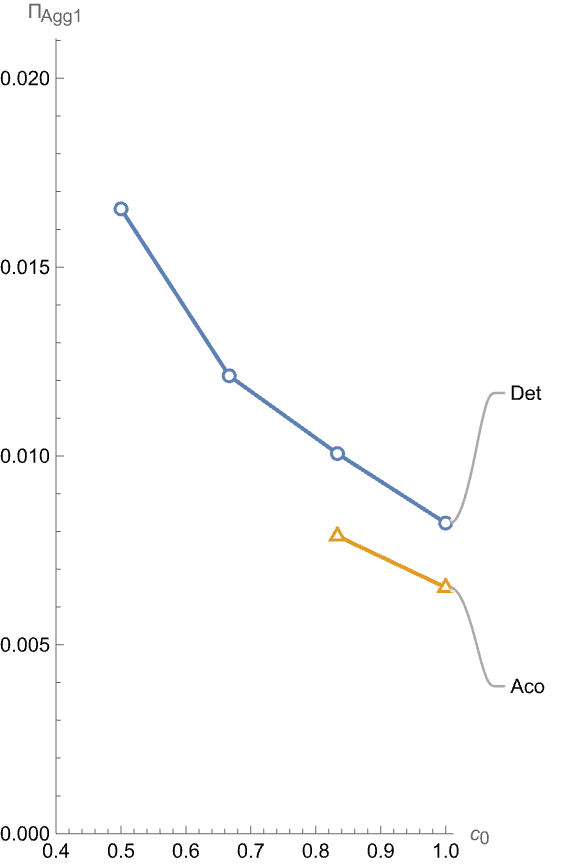}
\subcaption{Medium entry cost}\label{fig:ProfAgg1c0MedFPlot}
\end{subfigure}
\begin{subfigure}{.32\textwidth}
\includegraphics[width=\linewidth]{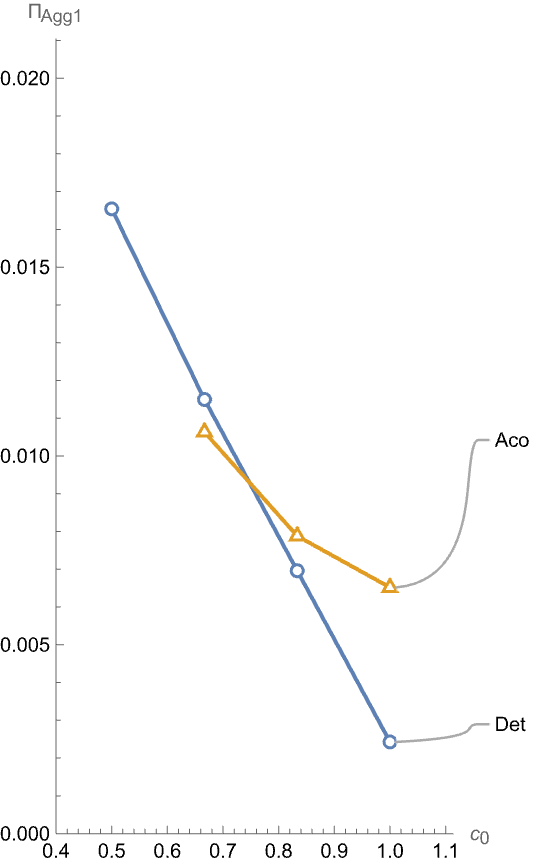}
\subcaption{Low entry cost}\label{fig:ProfAgg1c0LowFPlot}
\end{subfigure}
\caption{Maximum deterring and accommodating D1 profits as a function of $c_0$.}
\end{figure}

Fig.~\ref{fig:d0c0MedFPlot} shows, for medium entry cost, the profit maximizing $d_0$ when D1 correctly anticipates that D2 enters, so that D1 accomodates, or instead that D2 does not enter, so that D1 blockade/deters. As Fig.~\ref{fig:ProfAgg1c0MedFPlot} shows, blockade/deterrence is the only alternative for $c_0=3/6, 4/6$ and it is the preferred alternative for $c_0=5/6, 6/6$. Therefore, D1 either blockades or deters. In order to distinguish between blockade and deterrence, the solution to~\eqref{eq:D1deter_PMP} and the profit maximizing $d_0$ under monopsony are compared (\cref{fig:d0Detc0Plot}): when both values are equal, D1 actually blockades, which happens at $c_0=3/6, 4/6$, while when the latter is smaller, D1 deters, which happens at $c_0=5/6, 6/6$.

\begin{figure}
\centering
\begin{subfigure}{.32\textwidth}
\includegraphics[width=\linewidth]{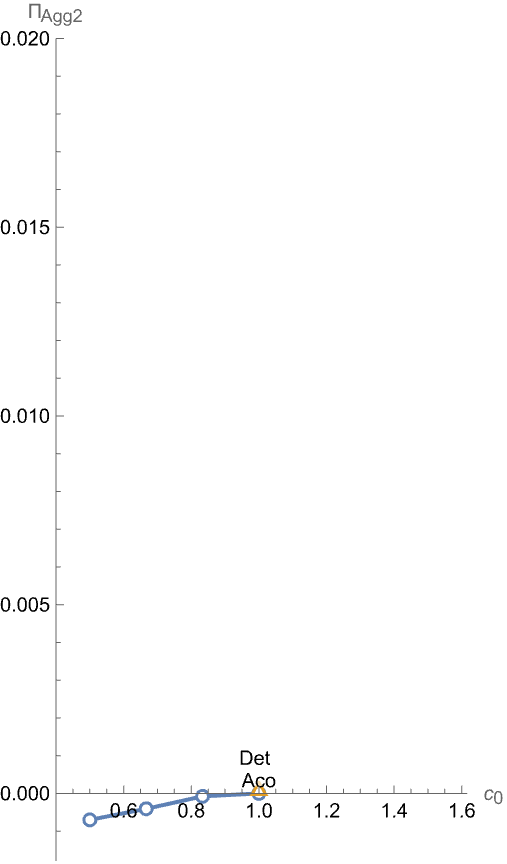}
\subcaption{High entry cost}\label{fig:ProfAgg2c0HighFPlot}
\end{subfigure}
\begin{subfigure}{.32\textwidth}
\includegraphics[width=\linewidth]{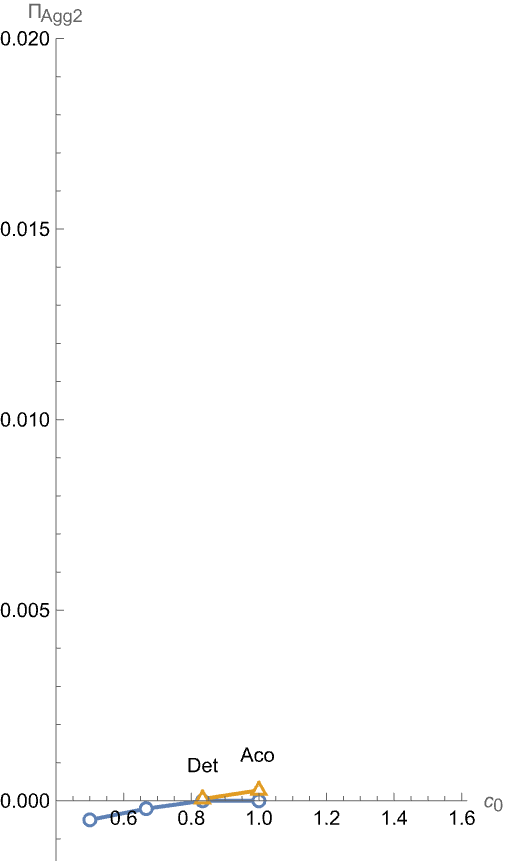}
\subcaption{Medium entry cost}\label{fig:ProfAgg2c0MedFPlot}
\end{subfigure}
\begin{subfigure}{.32\textwidth}
\includegraphics[width=\linewidth]{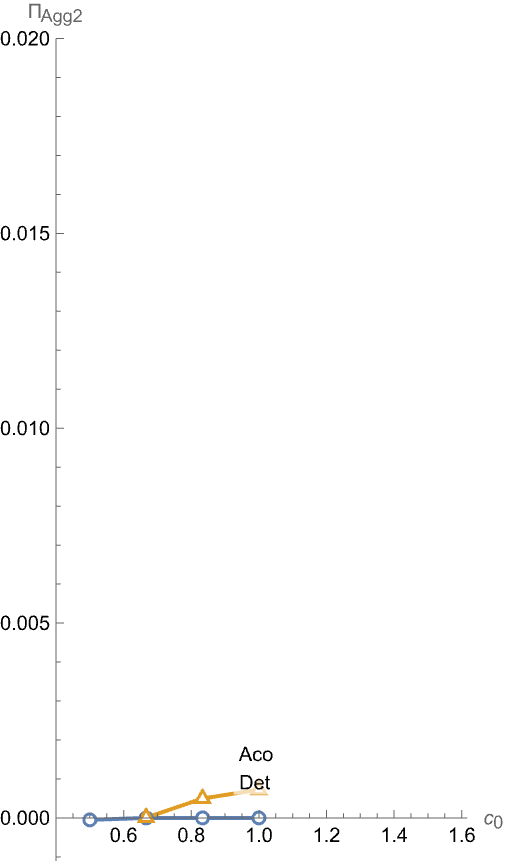}
\subcaption{Low entry cost}\label{fig:ProfAgg2c0LowFPlot}
\end{subfigure}
\caption{D2 Profit at deterring and accommodating $d_0$ as a function of $c_0$.}
\end{figure}

\begin{figure}
\centering
\begin{subfigure}{.32\textwidth}
\includegraphics[width=\linewidth]{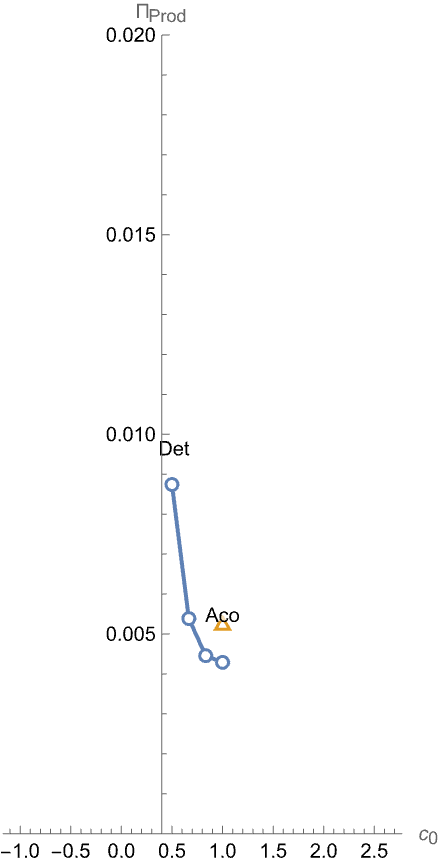}
\subcaption{High entry cost}\label{fig:ProfProdc0HighFPlot}
\end{subfigure}
\begin{subfigure}{.32\textwidth}
\includegraphics[width=\linewidth]{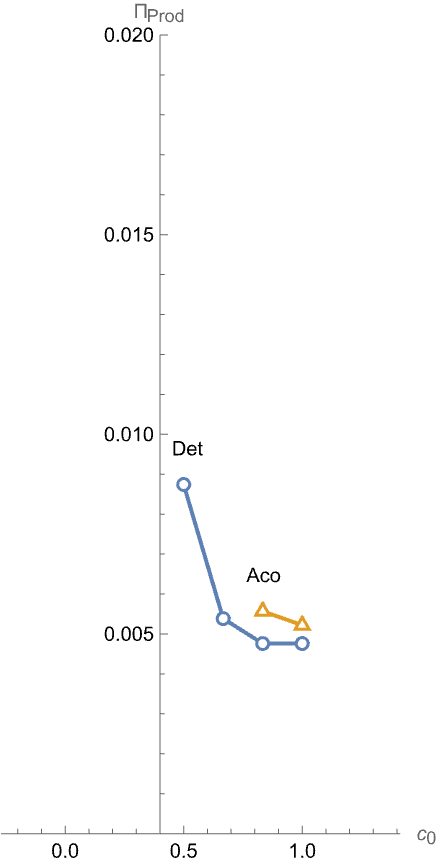}
\subcaption{Medium entry cost}\label{fig:ProfProdc0MedFPlot}
\end{subfigure}
\begin{subfigure}{.32\textwidth}
\includegraphics[width=\linewidth]{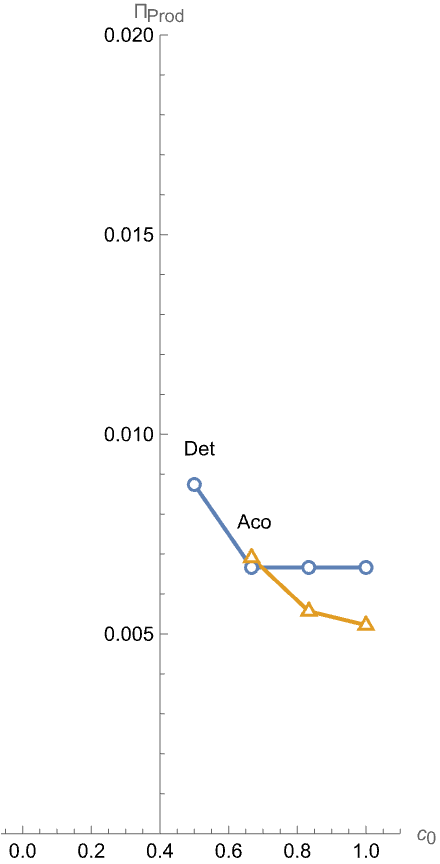}
\subcaption{Low entry cost}\label{fig:ProfProdc0LowFPlot}
\end{subfigure}
\caption{P1 Profit at deterring and accommodating $d_0$ as a function of $c_0$.}
\end{figure}

\begin{figure}
\centering
\begin{subfigure}{.32\textwidth}
\includegraphics[width=\linewidth]{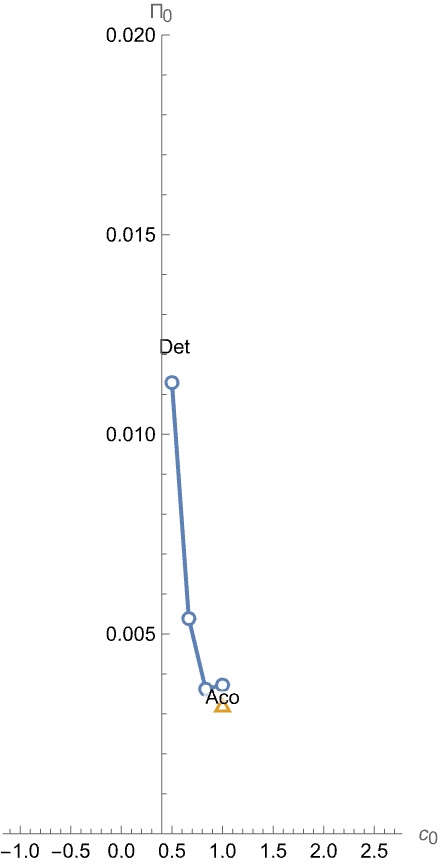}
\subcaption{High entry cost}\label{fig:ProfProd0c0HighFPlot}
\end{subfigure}
\begin{subfigure}{.32\textwidth}
\includegraphics[width=\linewidth]{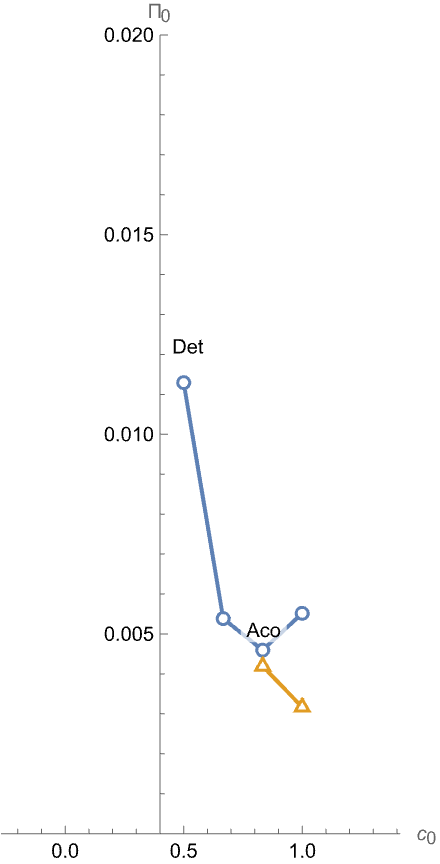}
\subcaption{Medium entry cost}\label{fig:ProfProd0c0MedFPlot}
\end{subfigure}
\begin{subfigure}{.32\textwidth}
\includegraphics[width=\linewidth]{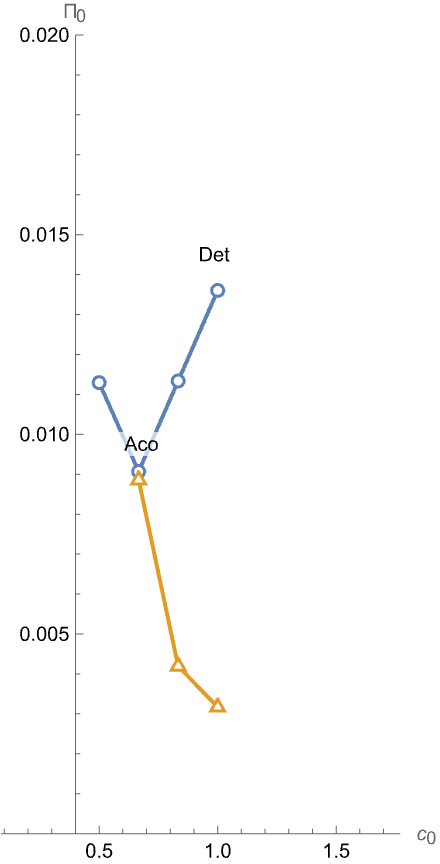}
\subcaption{Low entry cost}\label{fig:ProfProd0c0LowFPlot}
\end{subfigure}
\caption{P0 Profit at deterring and accommodating $d_0$ as a function of $c_0$.}
\end{figure}

\begin{figure}
\centering
\begin{subfigure}{.32\textwidth}
\includegraphics[width=\linewidth]{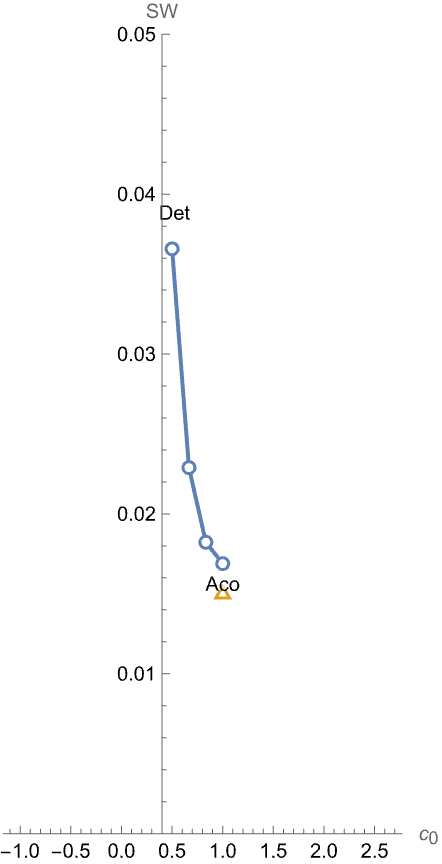}
\subcaption{High entry cost}\label{fig:SWc0HighFPlot}
\end{subfigure}
\begin{subfigure}{.32\textwidth}
\includegraphics[width=\linewidth]{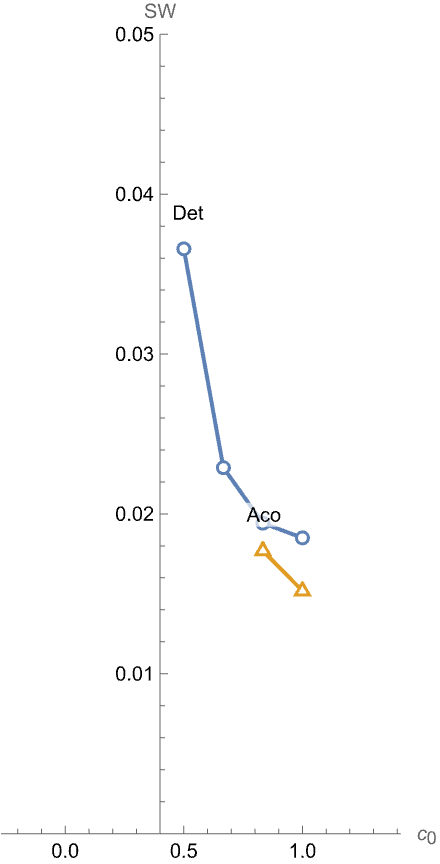}
\subcaption{Medium entry cost}\label{fig:SWc0MedFPlot}
\end{subfigure}
\begin{subfigure}{.32\textwidth}
\includegraphics[width=\linewidth]{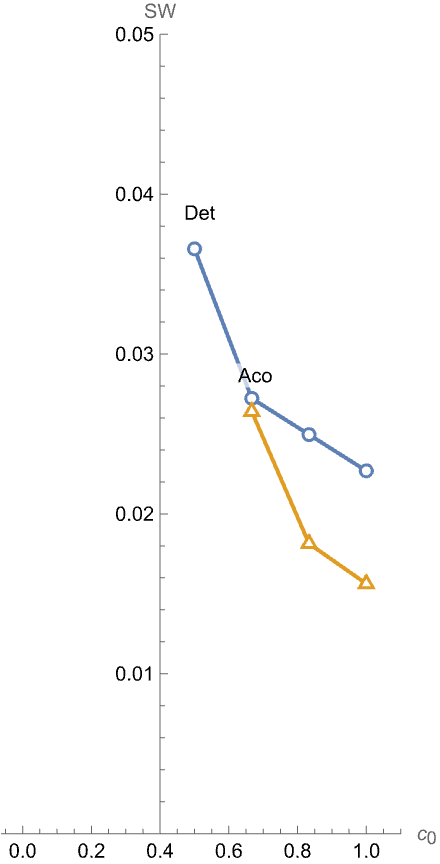}
\subcaption{Low entry cost}\label{fig:SWc0LowFPlot}
\end{subfigure}
\caption{Social welfare at deterring and accommodating $d_0$ as a function of $c_0$.}
\end{figure}

Aggregator D2 is, by definition, never worse off being accommodated than being deterred or blockaded (Fig.~\ref{fig:ProfAgg2c0MedFPlot}). Producer P1 is better off supplying both aggregators than just one, when feasible (Fig.~\ref{fig:ProfProdc0MedFPlot}). Producer P0, which is the exclusive supplier to D1, is nevertheless better off under blockade/deterrence, with aligned incentives to aggregator D1 (Fig.~\ref{fig:ProfProd0c0MedFPlot}). Finally, social welfare is higher under deterrence than under accommodation, when both outcomes are feasible (Fig.~\ref{fig:SWc0MedFPlot}). All profits, except D2 profits, decrease as $c_0$ increases, since less data is procured.

When entry costs are high, accommodation feasibility is reduced, as Fig.~\ref{fig:d0c0HighFPlot} shows. The discussion is the same as the medium entry cost case.

When entry costs are low, accommodation feasibility increases, as Fig.~\ref{fig:d0c0LowFPlot} shows. Now, as \cref{fig:ProfAgg1c0LowFPlot,fig:d0Detc0Plot} show, blockade/deterrence (blockade, actually) is the only alternative for $c_0=3/6$, deterrence is the preferred alternative for $c_0=4/6$, and accommodation is the preferred alternative for $c_0=5/6, 6/6$. Therefore, D1 either blockades, deters or accommodates as $c_0$ increases. The preferences for P0 and the whole are the same as previously discussed, while for P1 the preferences are reversed.

\FloatBarrier

\subsection{Comparative statics: \texorpdfstring{$\delta$}{delta}}

We vary $\delta$, which is the scope parameter, in the range $[ 0, 4 ]$, where $\delta=0$ denotes the absence of economies of scope.

\begin{figure}[!b]
\centering
\begin{subfigure}{.32\textwidth}
\includegraphics[width=\linewidth]{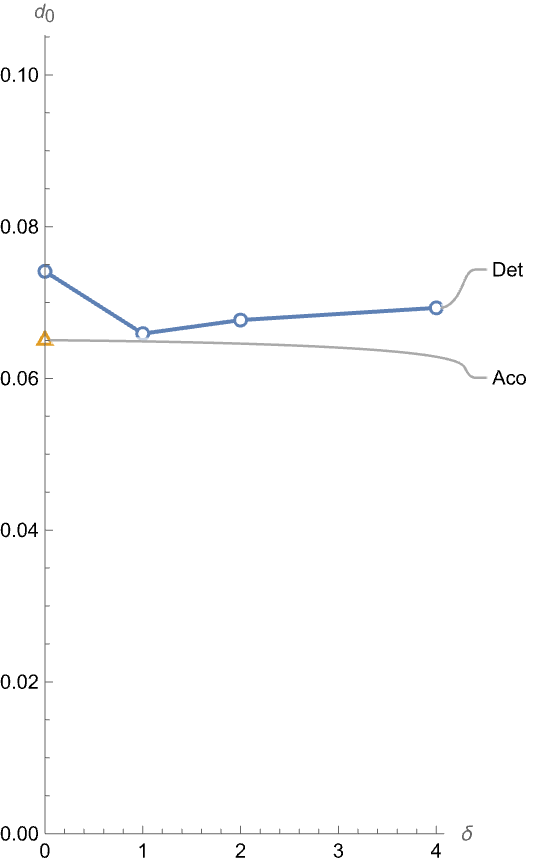}
\subcaption{High entry cost}\label{fig:d0deltaHighFPlot}
\end{subfigure}
\begin{subfigure}{.32\textwidth}
\includegraphics[width=\linewidth]{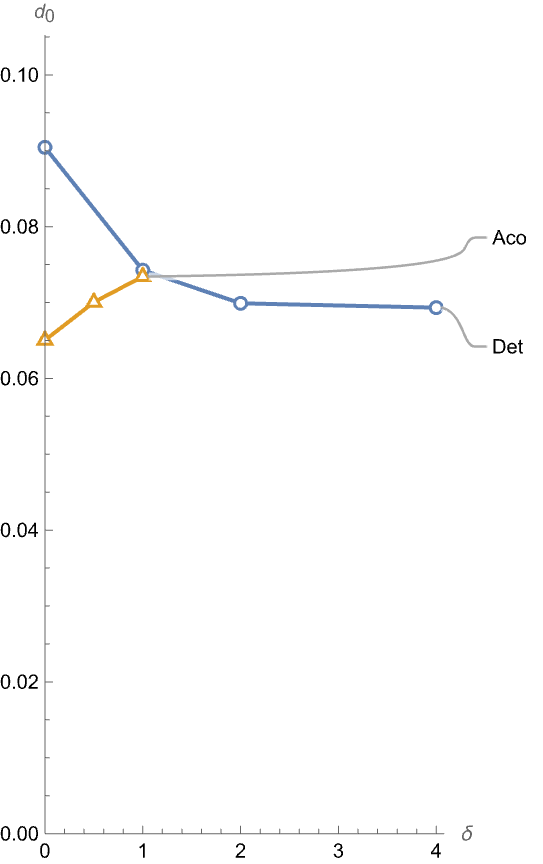}
\subcaption{Medium entry cost}\label{fig:d0deltaMedFPlot}
\end{subfigure}
\begin{subfigure}{.32\textwidth}
\includegraphics[width=\linewidth]{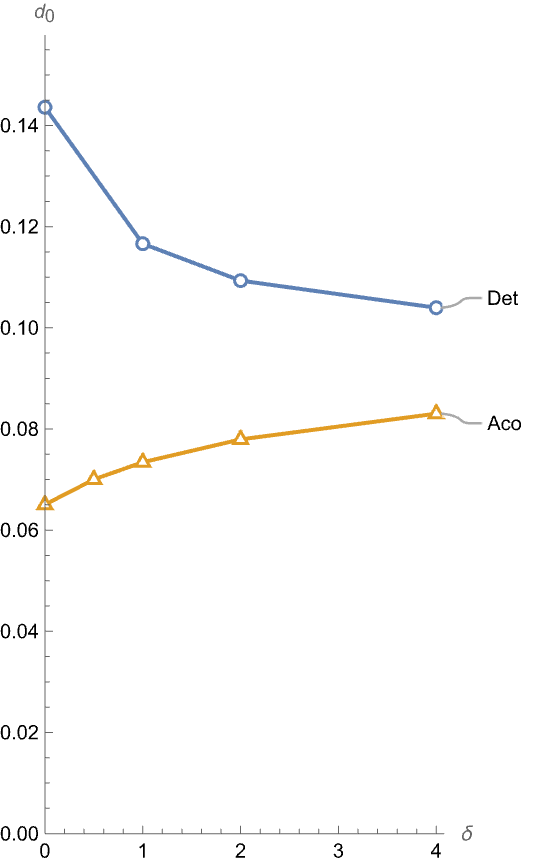}
\subcaption{Low entry cost}\label{fig:d0deltaLowFPlot}
\end{subfigure}
\caption{Profit maximizing deterring and accommodating $d_0$ as a function of $\delta$.}
\end{figure}

\begin{figure}
\centering
\begin{subfigure}{.32\textwidth}
\includegraphics[width=\linewidth]{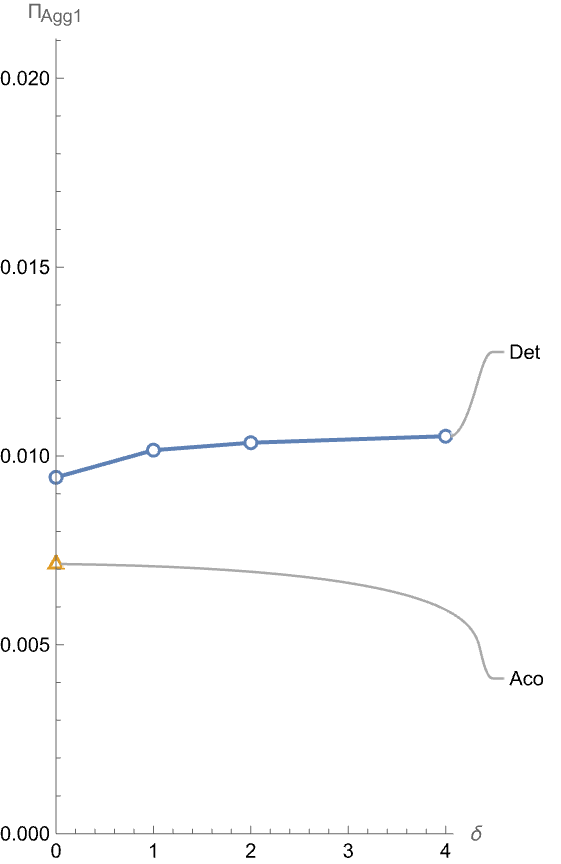}
\subcaption{High entry cost}\label{fig:ProfAgg1deltaHighFPlot}
\end{subfigure}
\begin{subfigure}{.32\textwidth}
\includegraphics[width=\linewidth]{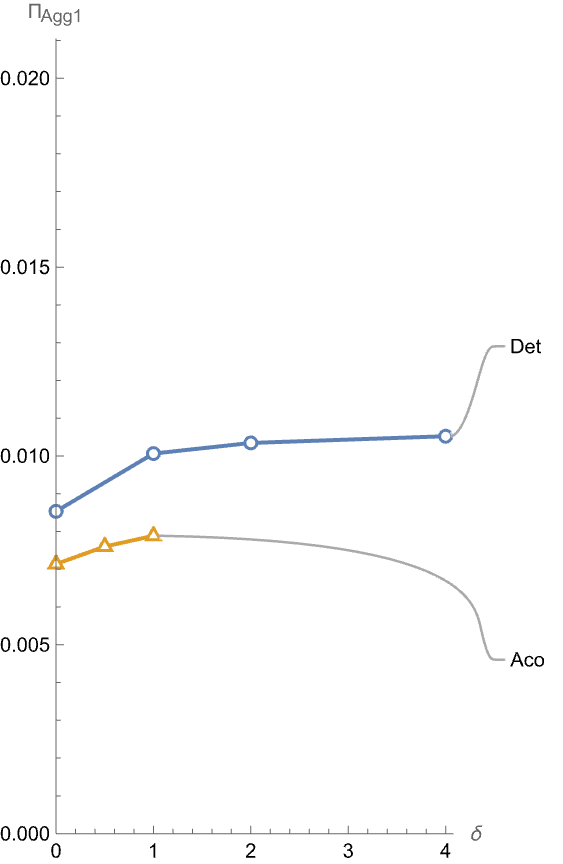}
\subcaption{Medium entry cost}\label{fig:ProfAgg1deltaMedFPlot}
\end{subfigure}
\begin{subfigure}{.32\textwidth}
\includegraphics[width=\linewidth]{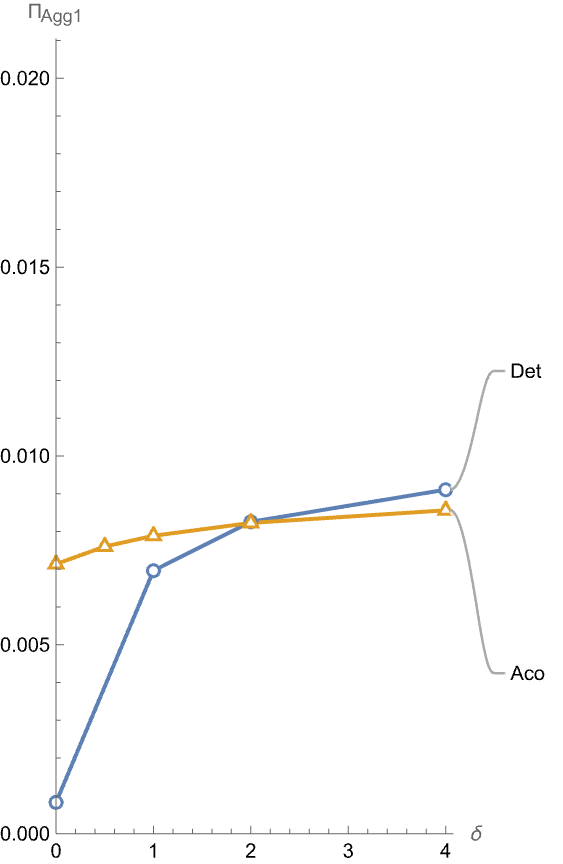}
\subcaption{Low entry cost}\label{fig:ProfAgg1deltaLowFPlot}
\end{subfigure}
\caption{Maximum deterring and accommodating D1 profits as a function of $\delta$.}
\end{figure}

Some discussion from previous section are the same, here. First, the effect of entry costs on the accommodation feasibility (\cref{fig:d0deltaMedFPlot,fig:d0deltaHighFPlot,fig:d0deltaLowFPlot}), whereby the lower the entry cost, the more feasible accommodation values exist, is the same. And second, the effect of entry costs on the incentive to accommodate (\cref{fig:ProfAgg1deltaMedFPlot,fig:ProfAgg1deltaHighFPlot,fig:ProfAgg1deltaLowFPlot}), whereby low entry costs compel the incumbent to accommodate, is also the same.

\begin{figure}
\centering
\begin{subfigure}{.32\textwidth}
\includegraphics[width=\linewidth]{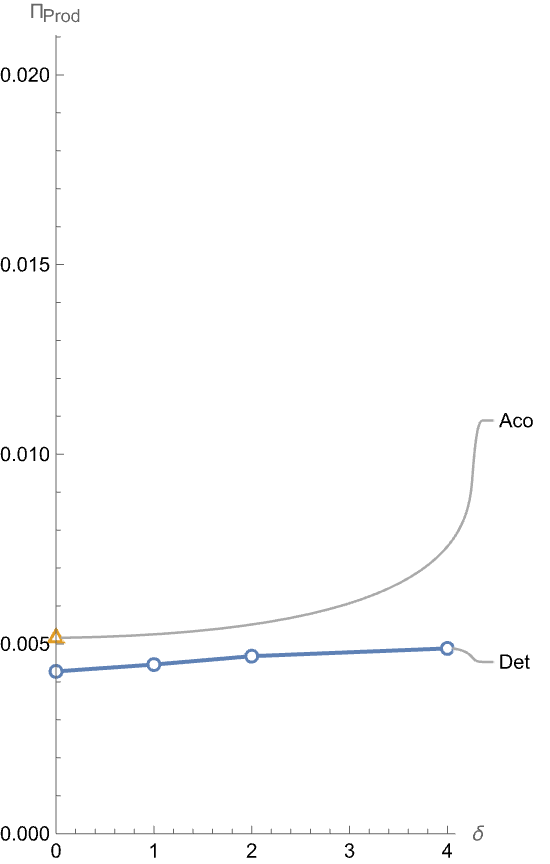}
\subcaption{High entry cost}\label{fig:ProfProddeltaHighFPlot}
\end{subfigure}
\begin{subfigure}{.32\textwidth}
\includegraphics[width=\linewidth]{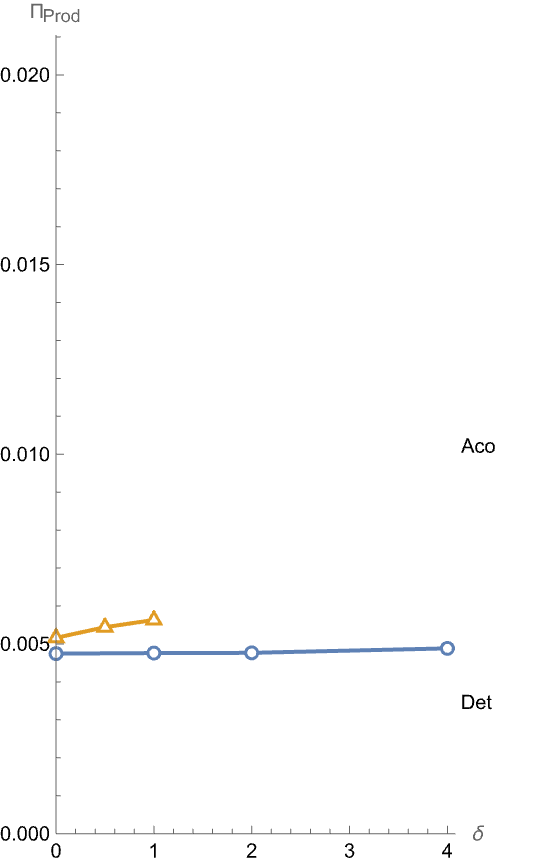}
\subcaption{Medium entry cost}\label{fig:ProfProddeltaMedFPlot}
\end{subfigure}
\begin{subfigure}{.32\textwidth}
\includegraphics[width=\linewidth]{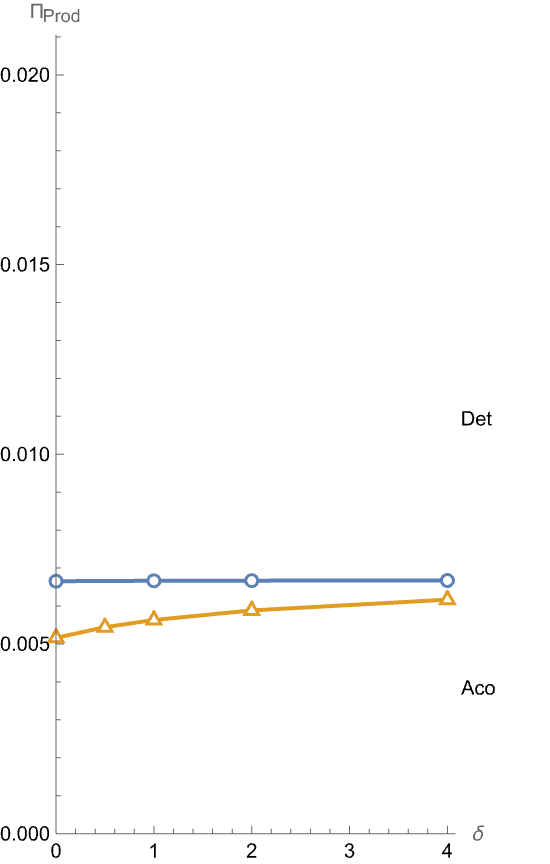}
\subcaption{Low entry cost}\label{fig:ProfProddeltaLowFPlot}
\end{subfigure}
\caption{P1 Profit at deterring and accommodating $d_0$ as a function of $\delta$.}
\end{figure}

\begin{figure}
\centering
\begin{subfigure}{.32\textwidth}
\includegraphics[width=\linewidth]{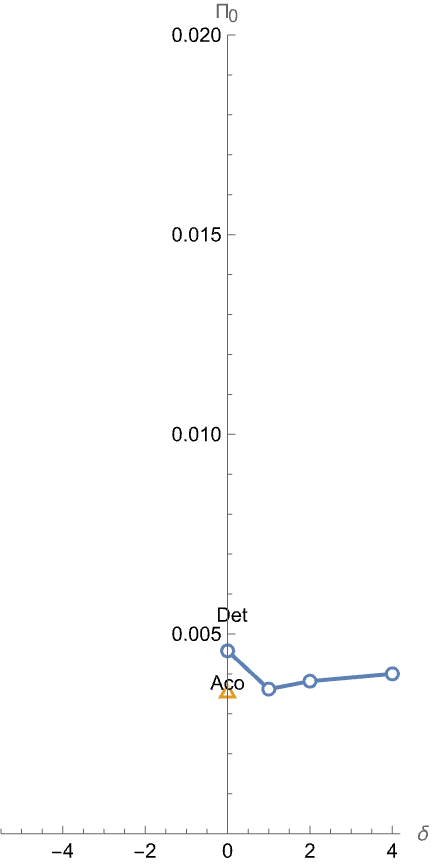}
\subcaption{High entry cost}\label{fig:ProfProd0deltaHighFPlot}
\end{subfigure}
\begin{subfigure}{.32\textwidth}
\includegraphics[width=\linewidth]{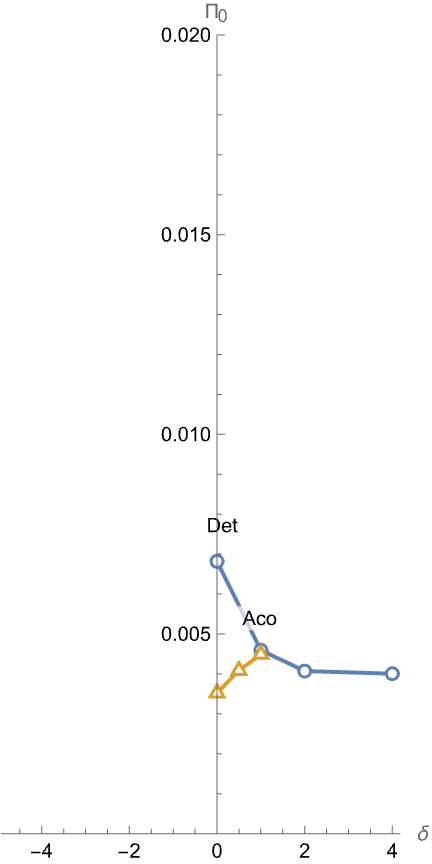}
\subcaption{Medium entry cost}\label{fig:ProfProd0deltaMedFPlot}
\end{subfigure}
\begin{subfigure}{.32\textwidth}
\includegraphics[width=\linewidth]{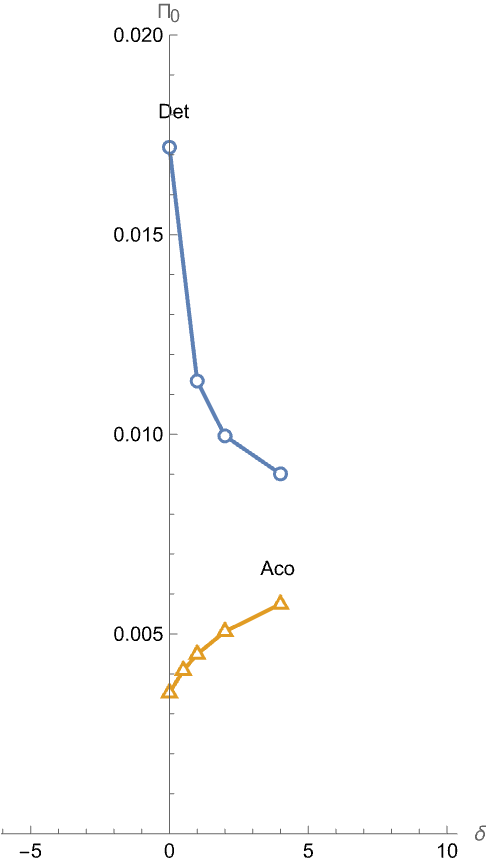}
\subcaption{Low entry cost}\label{fig:ProfProd0deltaLowFPlot}
\end{subfigure}
\caption{P0 Profit at deterring and accommodating $d_0$ as a function of $\delta$.}
\end{figure}

\begin{figure}
\centering
\begin{subfigure}{.32\textwidth}
\includegraphics[width=\linewidth]{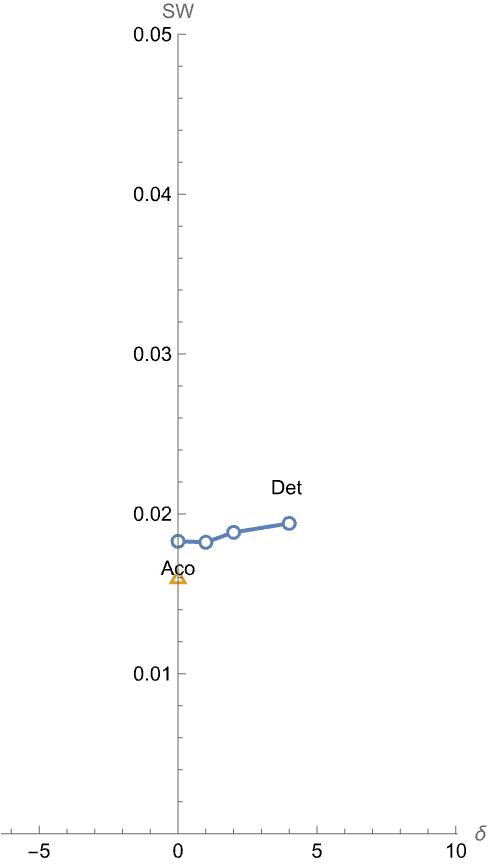}
\subcaption{High entry cost}\label{fig:SWdeltaHighFPlot}
\end{subfigure}
\begin{subfigure}{.32\textwidth}
\includegraphics[width=\linewidth]{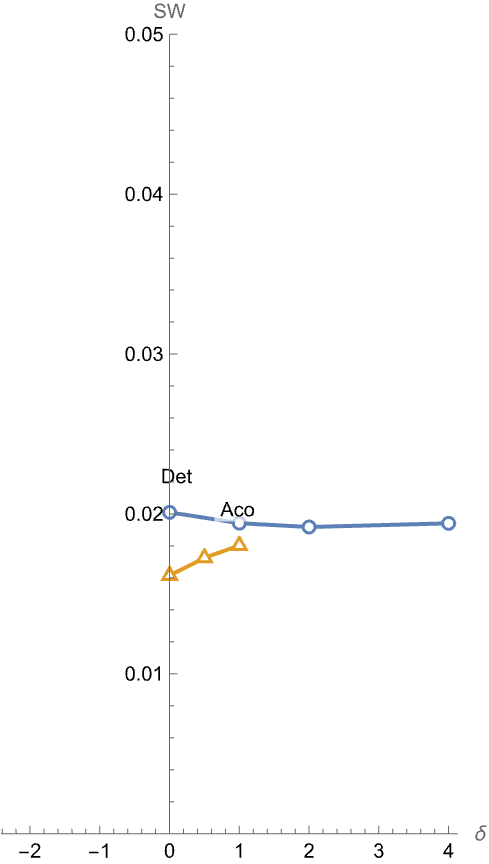}
\subcaption{Medium entry cost}\label{fig:SWdeltaMedFPlot}
\end{subfigure}
\begin{subfigure}{.32\textwidth}
\includegraphics[width=\linewidth]{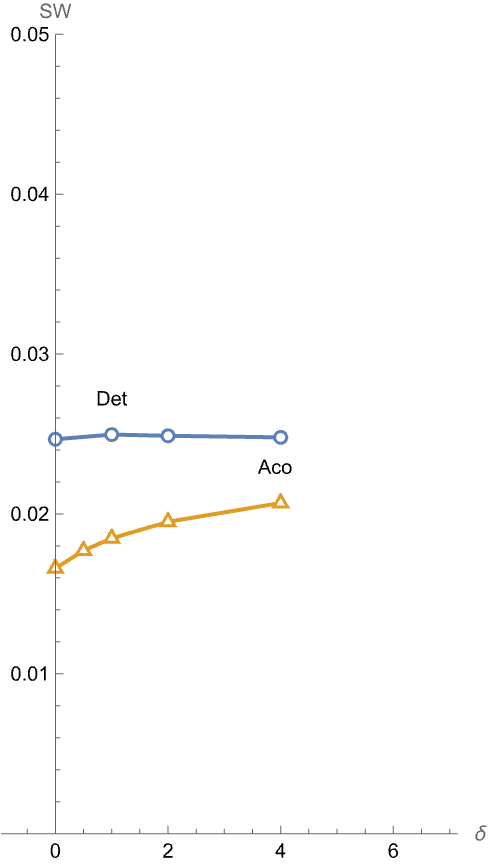}
\subcaption{Low entry cost}\label{fig:SWdeltaLowFPlot}
\end{subfigure}
\caption{Social welfare at deterring and accommodating $d_0$ as a function of $\delta$.}
\end{figure}

The effect of parameter $\delta$ on D1's, P1's and P0's profits and on social welfare (\cref{fig:ProfAgg1deltaMedFPlot,fig:ProfProddeltaMedFPlot,fig:ProfProd0deltaMedFPlot,fig:SWdeltaMedFPlot}, for medium entry cost) is similar to the effect of $c_0$, as long as an equivalence is made between a decrease in $c_0$ and an increase in $\delta$, since both facilitate the incumbent competitive advantage, either by reducing exclusive data procurement cost or by increasing the economies of scope, respectively. 

\begin{figure}[t]
\begin{center}
\includegraphics[width=\figwidth]{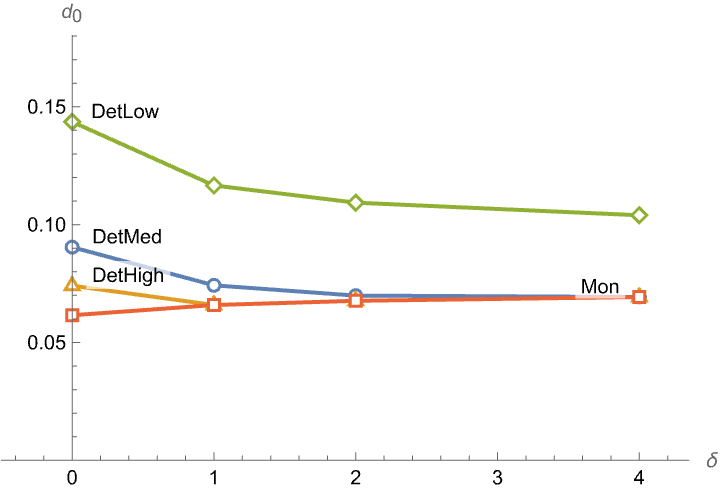}
\caption{Profit maximizing deterring and monopsonist $d_0$ as a function of $\delta$. Different entry costs.}\label{fig:d0DetdeltaPlot}
\end{center}
\end{figure} 

\begin{figure}
\centering
\begin{subfigure}{.32\textwidth}
\includegraphics[width=\linewidth]{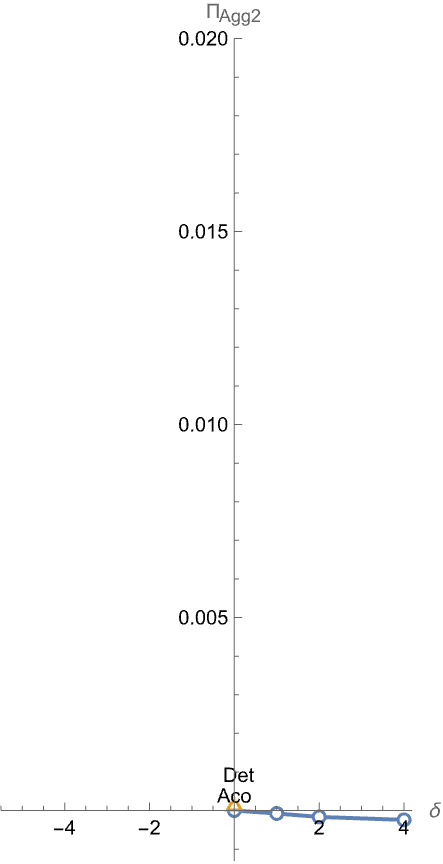}
\subcaption{High entry cost}\label{fig:ProfAgg2deltaHighFPlot}
\end{subfigure}
\begin{subfigure}{.32\textwidth}
\includegraphics[width=\linewidth]{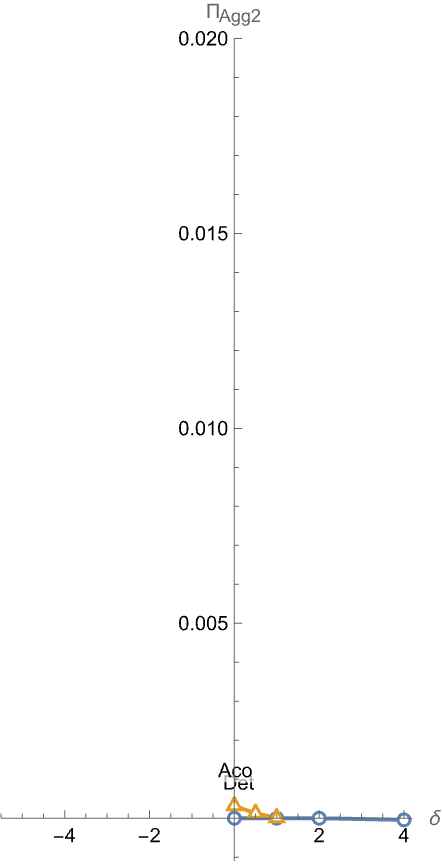}
\subcaption{Medium entry cost}\label{fig:ProfAgg2deltaMedFPlot}
\end{subfigure}
\begin{subfigure}{.32\textwidth}
\includegraphics[width=\linewidth]{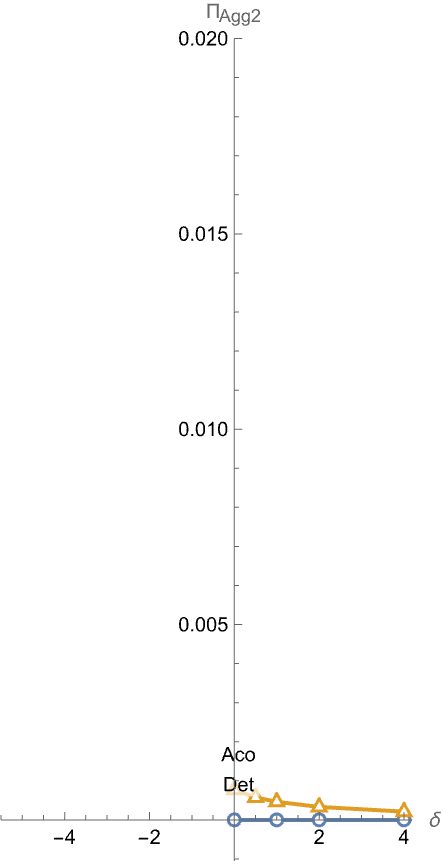}
\subcaption{Low entry cost}\label{fig:ProfAgg2deltaLowFPlot}
\end{subfigure}
\caption{D2 Profit at deterring and accommodating $d_0$ as a function of $\delta$.}
\end{figure} 

\FloatBarrier

\section{Conclusions}\label{sec:conclusions}
 
In this work, we have modeled a market for data where an incumbent and a challenger compete for data from a producer. The incumbent has access to an exclusive data producer, and it uses this exclusive access, together with economies of scope in the aggregation of the data, as a strategy against the potential entry by the challenger. We have assessed the incumbent incentives to either deter or accommodate the entry of the challenger.

We have shown that the incumbent will accommodate (i.e, it is in the incumbent's interest to accommodate) when the exclusive access is costly and when the economies of scope are low, and it will blockade or deter otherwise. Although blockade/deterrence is more efficient (i.e., the social welfare is higher) than accommodation, this result is linked to the fact that the model is limited to the data aggregators and the data producers. If the model had encompassed also a market for a service supplied by the data aggregators, then a service monopoly (i.e., served only by D1) would have resulted in a lower total welfare than a service duopoly (i.e., served by both D1 and D2). 

The above results and discussion would justify an access regulation that incentivizes the entry of the challenger, e.g., by increasing production costs for the exclusive data. Other regulatory measures, such as non-discriminatory access to the exclusive data producer are outside the scope of this work. 

\section*{Acknowledgment} 
Financial support is acknowledged from Grant PID2021-123168NB-I00, funded by MCIN/AEI, Spain/10.13039/ 501100011033 and the European Union A way of making Europe/ERDF and Grant TED2021-131387B-I00, funded by MCIN/AEI, Spain/ 10.13039/501100011033 and the European Union NextGenerationEU/ RTRP.

\bibliography{bib}

\end{document}